\documentclass[preprint,aps,tightenlines,showpacs,nofootinbib]{revtex4}
\usepackage{latexsym}
\usepackage[dvips]{graphicx}
\newcommand{\beq}{\begin{eqnarray}}
\newcommand{\eeq}{\end{eqnarray}}
\newcommand{\eq}{eqnarray}

\newcommand{\al}{{\alpha}}
\newcommand{\be}{{\beta}}
\newcommand{\ci}{\cite}
\newcommand{\ga}{{\gamma}}

\newcommand{\ep}{{\epsilon}}

\newcommand{\de}{{\delta}}

\newcommand{\la}{{\lambda}}
\newcommand{\La}{{\Lambda}}
\newcommand{\m}{{\mu}}
\newcommand{\n}{{\nu}}
\newcommand{\ka}{{\kappa}}

\newcommand{\om}{{\omega}}
\newcommand{\Om}{{\Omega}}
\newcommand{\pa}{{\partial}}
\newcommand{\no}{{\nonumber}}
\newcommand{\f}{\frac}
\newcommand{\ra}{\rightarrow}

\begin{document}

\preprint{arXiv:0910.5117v4 [hep-th]}

\title{Ho\v{r}ava Gravity and Gravitons 
at a Conformal Point}

\author{Mu-In Park\footnote{E-mail address: muinpark@gmail.com}}

\affiliation{
The Institute of Basic Sciences, Kunsan National University, Kunsan,
573-701, Korea\footnote{Past Address: Research Institute of Physics
and Chemistry, Chonbuk National University, Chonju 561-756, Korea }
}

\begin{abstract}
Recently Ho\v{r}ava proposed a renormalizable gravity theory with
higher derivatives by abandoning the Lorentz invariance in UV. Here,
I 
construct the Ho\v{r}ava model at $\la=1/3$, where a local
anisotropic Weyl symmetry exists in the UV limit, in addition to the
foliation-preserving diffeomorphism. By considering linear
perturbations around Minkowski vacuum for the non-projectable
version of the Ho\v{r}ava model, I show that the scalar graviton
mode is completely disappeared and only the usual tensor graviton
modes remain in the physical spectrum. The existence of the UV
conformal symmetry is unique to the theory with the detailed balance
and this may explain the importance of the detailed balance
condition in quantum gravity.

\end{abstract}


\maketitle

\newpage

\section{Introduction}
Recently Ho\v{r}ava proposed a renormalizable gravity theory with
higher spatial derivatives (up to sixth order) in four dimensions
which reduces to Einstein gravity with a {\it non-vanishing}
cosmological constant in IR but with improved UV behaviors, by
abandoning the Lorentz invariance from non-equal-footing treatment
of space and time \ci{Hora:08,Hora}.  Due to lack of the full
diffeomorphism ({\it Diff}), some extra graviton modes are expected
generally but there have been confusions regarding the extra modes
and the consistency of the Ho\v{r}ava model
\ci{Cai:0905,Keha,Chen:0905_2,Char,Li,Soti,Kim,Muko:0905,Gao:0905,
Blas:0906,Koba:0906,Bogd:0907,Wang:0907}. But, recently I have
reconsidered these problems for the non-projectable version of the
Ho\v{r}ava model and showed that, in the Minkowski vacuum
background, the extra scalar graviton mode can be consistently
decoupled from the usual tensor graviton modes, by imposing the
(local) Hamiltonian constraint as well as the momentum constraints
\ci{Park:0910}, for arbitrary values of the Lorentz breaking
parameter $\la$, except the point $\la= 1/3$, where the theory
becomes singular and needs a separate consideration. This reduces to
the results of Einstein gravity in IR and achieves the consistency
of the model.

In this paper, I consider the Ho\v{r}ava model at the particular
point $\la=1/3$, where local anisotropic Weyl rescaling symmetry
exists in the UV limit, in addition to the foliation-preserving {\it
Diff}. Due to the existence of the extra (local) symmetry in UV, it
was demonstrated that there would be {\it no} physical excitation of
the extra scalar mode in UV \ci{Hora}. But this analysis was not
enough to understand the complete aspects of the system. For
example, the Cotton-squared term was crucial for the scalar
graviton's UV decoupling but this peculiar term can only be
naturally explained by the detailed balance condition, though its
physical meaning was unclear in the non-conformal cases. And also,
the analysis was only for the UV limit and so more extension of the
analysis was needed to see the complete spectrum of all the graviton
modes. Finally, due to the ``local'' symmetry in UV, it has been
noted that the lapse function $N$ can no longer be a projectable
function on the foliation, which means $N=N(t)$ for the foliation
time $t$, in the conformal case, in contrast to non-conformal cases
\ci{Hora}. So, it would be important to see the consequences of the
non-projectability in the complete spectrum of graviton modes at the
conformal point $\la=1/3$. So, the object of this paper is twofold.
First is, for the first time, the construction of the full Horava
action at the conformal point, which provides a natural explanation
of the appearance of Cotton-squared term from the detailed balance
condition as well as other lower derivatives terms. Second is the
complete analysis of the linear perturbation for the particular
system in parallel with non-conformal case of Refs.
\ci{Keha,Park:0910}, where the lapse function $N$ is considered as a
non-projectable function on the foliation, i.e., a function of space
${\bf x}$ as well as $t$, $N=N({\bf x},t)$.

The plan of this paper is as follows. In Sec. II, I first construct
the Ho\v{r}ava model at the conformal point $\la=1/3$, which needs a
separate consideration due to degeneracy of the DeWitt metric, based
on the standard framework with the detailed balance condition
\ci{Hora:08,Hora,Park:0905,Park:0906}. In Sec. III, I consider the
non-projectable version of the model and show that the scalar
graviton mode is completely disappeared in the physical spectrum.
Only the usual tensor graviton modes remain, as expected, by
considering linear perturbations of metric around Minkowski vacuum
and imposing the (local) Hamiltonian constraint, as well as the
momentum constraints, from the non-projectable lapse function
\ci{Park:0910}. In Sec. IV, I conclude with several discussions.


\section{Ho\v{r}ava Gravity at a Conformal Point}

Ho\v{r}ava gravity is defined as a power-counting renormalizable
gravity, by introducing the anisotropic UV scaling
\begin{\eq}
\label{rigid} {\bf x} \ra s {\bf x},~t \ra s^z t
\end{\eq}
 with $z=3$ in (3+1) spacetime
dimensions \ci{Hora}. By considering the ADM decomposition of the
metric
\begin{\eq}
ds^2=-N^2 c^2 dt^2+g_{ij}\left(dx^i+N^i dt\right)\left(dx^j+N^j
dt\right)\ ,
\end{\eq}
the Ho\v{r}ava action reads, formally,
\begin{\eq}
\label{Horava_def}
&&S=\int dt d^3 x\sqrt{g}N
\left\{\frac{2}{\kappa^2}
K_{ij}G^{ijkl}K_{kl}-\frac{\kappa^2}{2}\left[\frac{1}{\n^2}C^{ij}
-\frac{\mu}{2}\left(R^{(3)ij}-\frac{1}{2}R^{(3)}g^{ij}+\La_W
g^{ij}\right)\right] \right.\cr
&&\qquad\qquad\qquad\qquad\left.{}\times{\cal G}_{ijkl}
\left[\frac{1}{\n^2}C^{kl}-\frac{\mu}{2}\left(R^{ (3)
k\ell}-\frac{1}{2}R^{(3)}g^{kl} +\La_W g^{kl}\right)\right]\right\},
\end{\eq}
where
\begin{\eq}
 K_{ij}=\frac{1}{2N}\left(\dot{g}_{ij}-\nabla_i
N_j-\nabla_jN_i\right)\
 \end{\eq}
is the extrinsic curvature (the dot $(\dot{~})$ denotes the
derivative with respect to $t$),
\begin{\eq}
 C^{ij}=\epsilon^{ik\ell}\nabla_k
\left(R^{(3)j}{}_\ell-\frac{1}{4}R^{(3)} \delta^j_\ell\right)\
 \end{\eq}
is the Cotton tensor,  $\kappa,\lambda,\nu,\mu$, and $\La_W$ are
constant parameters. The supermetric
\begin{\eq}
G^{ijkl}=\de^{ijkl}-\lambda g^{ij}g^{kl},
\end{\eq}
with $\de^{ijkl} \equiv
\frac{1}{2}(g^{ik}g^{j\ell}+g^{i\ell}g^{jk})$, is the generalized
DeWitt metric for a Lorentz symmetry breaking parameter $\la$, which
is $1$ for the usual Lorentz invariant general relativity (GR) and
its deviation from $1$ measures the violation of Lorentz symmetry in
the kinetic term,
\begin{\eq}
S_K=\int dt d^3 x\sqrt{g}N \frac{2}{\kappa^2} K_{ij}G^{ijkl}K_{kl} .
\end{\eq}
Another supermetric ${\cal G}_{ijkl}$ is the inverse of the DeWitt
metric, satisfying $G^{ijmn}{\cal G}_{mnkl}=\de^{ij}_{kl}$. And in
deriving all the remaining terms, other than the kinetic terms, in
(\ref{Horava_def}), which is called as the potential terms, I have
adopted the ``detailed balance" prescription as
\begin{\eq}
S_V=-\frac{\kappa^2}{8} \int dt d^3 x \sqrt{g}N \f{\de W}{\de
g^{ij}} {\cal G}_{ijkl} \f{\de W}{\de g^{kl}}
\end{\eq}
with
\begin{\eq}
\f{\de W}{2 \de g^{ij}}=\frac{1}{\n^2}C^{ij}
-\frac{\mu}{2}\left(R^{(3)ij}-\frac{1}{2}R^{(3)}g^{ij}+\La_W
g^{ij}\right)
\end{\eq}
from the three-dimensional ({\it Euclidean}) gravity action
\ci{Witt,Park:0705}
\begin{\eq}
W=\frac{1}{\nu^2}\int Tr \left(\Gamma\wedge d\Gamma+\frac{2}{3}
\Gamma\wedge\Gamma \wedge\Gamma \right)+\mu \int d^3 x
\sqrt{g}(R^{(3)}-2\La_W).
\end{\eq}

For $\la \neq 1/3$ ($1/D$ in $D$ spatial dimensions), the DeWitt
metric is not degenerated and the explicit form of the action can be
obtained by considering the inverse DeWitt metric ${\cal
G}_{ijkl}=\de_{ijkl}-\f{\la}{3 \la-1} g_{ij}g_{kl}$ with $\de_{ijkl}
\equiv \frac{1}{2}(g_{ik}g_{j\ell}+g_{i\ell}g_{jk})$
\ci{Hora:08,Hora}. On the other hand, for $\la=1/3$, the DeWitt
metric is degenerated and we need to project out the non-degenerate
parts only when considering the inverse DeWitt metric. Actually, by
using the fact that $G^{ijkl}$ has a null eigenvector $g_{ij}$,
\begin{\eq}
G^{ijkl} g_{ij}=0,
\end{\eq}
it is easy to see that its inverse ${\cal G}_{ijkl}$ is given by
\begin{\eq}
{\cal G}_{ijkl}&=&\de_{ijkl}-\f{1}{3} g_{ij}g_{kl}, \no \\
{\cal G}_{ijkl} g^{ij} &=&0,~{\cal G}_{ijmn}
G^{mnkl}={\tilde{\de}_{ij}}^{kl}
\end{\eq}
with the (projected) Kronecker-delta
${\tilde{\de}_{ij}}^{kl}=\de_{ij}^{kl}-\f{1}{3}g_{ij} g^{kl}$,
satisfying ${\tilde{\de}_{ij}}^{kl}g^{ij}=0$.

Then, after some manipulations, one can find the following action,
from (\ref{Horava_def}),
\begin{\eq}
S &= & \int dt d^3 x \sqrt{g}N
\left[\frac{2}{\kappa^2}\left(K_{ij}K^{ij}-\f{1}{3}
K^2\right)-\frac{\kappa^2}{2\nu^4}C_{ij}C^{ij}+\frac{\kappa^2
\mu}{2\nu^2}\epsilon^{ijk} R^{(3)}_{i\ell} \nabla_{j}R^{(3)\ell}{}_k
\right. \nonumber \\ &&\left. -\frac{\kappa^2\mu^2}{8} \left(
R^{(3)}_{ij} R^{(3)ij}-\f{1}{3} (R^{(3)})^2 \right) \right] .
\label{Park}
\end{\eq}
Note that all terms which are proportional to $\La_W$ are canceled,
in sharply contrast to $\la \neq 1/3$ cases, and consequently there
is {\it no} term, proportional to $R^{(3)}$ and cosmological
constant term $\La$ of the usual GR. This is basically due to the
``detailed balance" condition and we need ``soft" breakings of the
detailed balance in order that the usual GR limit may be obtained in
IR\footnote{The recovery of GR would require $\lambda$ to flow from
$1/3$ in UV to $1$ in IR. But this is problematic in the, so-called,
``projectable'' version of the Ho\v{r}ava gravity, where the scalar
graviton exists and it becomes a ghost for $1/3< \lambda <1$
\ci{Bogd:0907} (see also \ci{Gong:1002}), i.e., in the process of
RG-flows.}. Then the desired form of the general action would be,
\begin{\eq}
\label{SIR}
 S_g &=&S + \int dt d^3 x \sqrt{g}N \frac{\kappa^2 \mu^2 \hat{\om}}{8} \left(
R^{(3)}-\f{2 \La}{c^2} \right)
\end{\eq}
by introducing the soft breaking terms of the detailed balance,
$R^{(3)},\La$ and these modify the IR behaviors
\ci{Hora,Keha,Park:0905,Park:0906}.

This action breaks the Lorentz symmetry manifestly even in IR limit
where the kinetic term and the last two IR-modification terms in
(\ref{SIR}) dominate, due to the Lorentz non-invariant deformation
of the kinetic term with $\la=1/3$. However, this action has another
symmetry, called ``anisotropic" Weyl rescaling symmetry,
\begin{\eq}
\label{Weyl}
 g_{ij} \ra \Om^2 (t, {\bf x}) ~g_{ij} ,~N  \ra \Om^z
(t, {\bf x}) ~N ,~N_i  \ra \Om^2 (t, {\bf x})~N_i
\end{\eq}
at each spacetime point: Under this transformations, the measure
part is transformed as $\sqrt{g} N \ra \Om^{3+z} \sqrt{g} N$ and,
for $z=3$, the $\Om^{3+z}=\Om^6$ factor is canceled by the
$\Om^{-6}$ factors from the kinetic parts and the highest (6th
order) spatial-derivative term $C_{ij}C^{ij}$. But this symmetry
exists only in the UV limit since all other lower-spatial derivative
terms violate this symmetry, explicitly; it is interesting to note
that the last term in (\ref{Park}) is, up to boundary terms, the
spatial part of the square of the (four-dimensional) Weyl tensor
$C_{\m \n \sigma \rho} C^{\m \n \sigma \rho}=\f{1}{2}(R_{\m \n}
R^{\m \n} -\f{1}{3} R^2)$ in the conformal gravity, which is
invariant under (\ref{Weyl}) for $z=1$ but not invariant for $z=3$.
So this is an ``emergent" symmetry in the UV limit only and this is
in contrast to the Lorentz symmetry, which emerges in the IR limit
for $\la=1$ \ci{Hora,Park:0910}. The existence of the UV conformal
symmetry is unique to the theory with the detailed balance: If one
does not adopt the detailed balance, one does not have the 6th-order
spatial-derivatives term $C_{ij}C^{ij}$ and the UV symmetry is
violated by the potential terms $S_V$, generally \ci{Soti}.
Actually, the conformal symmetry is {\it inherent} in the defining
properties of spacetimes with the scaling (\ref{rigid}) since the
transformation (\ref{Weyl}) corresponds to the local version of the
rigid anisotropic scaling (\ref{rigid}) with a constant $\Om=s$
\ci{Hora:08,Hora,Hora:0909}; the detailed balance is a natural way
to introduce the desired potential $C_{ij}C^{ij}$ into the action.

By comparing the IR limit of the general action (\ref{SIR}) with
$\la$-deformed Einstein-Hilbert action $S_{\la \rm EH}$
\ci{Park:0906},
\begin{\eq}
S_{\la \rm EH} &= & \f{c^4}{16 \pi G } \int dt d^3 x
\sqrt{g}N\left[\frac{1}{c^2}\left(K_{ij}K^{ij}-\lambda
K^2\right)+R^{(3)}-\f{2 \Lambda}{c^2} \right]\ \label{lEH}
\end{\eq}
one may obtain the fundamental constants of the speed of light $c$,
the Newton's constant $G$ as
\begin{\eq}
\label{constant}
 c=\sqrt{\f{\kappa^4 \mu^2
\hat{\om}}{16}},~G=\f{\kappa^2 c^2}{32 \pi}.
\end{\eq}

In the canonical formulation, the existence of the Weyl symmetry is
reflected in the primary constraint
\begin{\eq}
{\pi^i}_i \equiv g_{ij} \pi^{ij} \approx 0
\end{\eq}
for the momenta
\begin{\eq}
\pi^{ij} \equiv \f{\de S}{\de \dot{g}_{ij}} =\f{2
\sqrt{g}}{\kappa^2} G^{ijkl}K_{kl},
\end{\eq}
in addition to the usual Hamiltonian and momentum constraints. And
the symmetry in UV limit implies that the constraint leads to the
first-class constraints system in UV.

\section{Graviton Modes}

The action is invariant under the foliation-preserving {\it Diff},
\begin{\eq}
\label{Diff}
\delta x^i &=&-\zeta^i (t, {\bf x}), ~\delta t=-f(t), \no \\
 \delta
g_{ij}&=&\pa_i\zeta^k g_{jk}+\pa_j \zeta^k g_{ik}+\zeta^k
\pa_k g_{ij}+f \dot g_{ij},\nonumber\\
\delta N_i &=& \pa_i \zeta^j N_j+\zeta^j \pa_j N_i+\dot\zeta^j
g_{ij}+f \dot N_i+\dot f N_i,\no \\
\delta N&=& \zeta^j \pa_j N+f \dot N+\dot f N.
\end{\eq}
Note that this {\it Diff} exists for arbitrary spacetime-dependent
$N,N_i,g_{ij}$. This implies that the equations of motion by varying
$N,N_i,g_{ij}$ are all the ``local'' equations as in the usual
Lorentz invariant Einstein gravity (see \ci{Park:0910} for the
detailed discussions). Generally, it seems that there are two gauge
inequivalent classes of Ho\v{r}ava gravity, i.e, projectable and
non-projectable versions, depending on whether $N$ is a function of
$t$ only or not. However, from the recovery of GR in IR, which could
be problematic in the projectable version, as mentioned in the
footnote No.1, I only consider the non-projectable version in this
paper. This is in the same spirit as in the previous work on
$\lambda \neq 1/3$.

In order to study graviton modes, I need to consider perturbations
of metric around some appropriate backgrounds, which are solutions
of the full theory (\ref{SIR}). Here, I consider only the
perturbations around Minkowski vacuum, which is a solution of the
full theory (\ref{SIR}) in the limit of $\La \ra 0$, for simplicity,
\begin{\eq}
\label{pert}
 g_{ij}=\de_{ij} +\ep h_{ij}, ~ N=1+ \ep n, ~ N_i = \ep
n_i
\end{\eq}
with a small expansion parameter $\ep$.

From the extrinsic curvatures under the perturbations (\ref{pert}),
\begin{\eq}
K_{ij}&=&\f{\ep}{2} \left(\dot{h}_{ij}- \pa_i n_j- \pa_j n_i \right)
+ {\cal O}
(\ep^2), \no \\
K&=& \f{\ep}{2} \left(\dot{h}-2 \pa_i n^i \right) + {\cal O} (\ep^2)
\end{\eq}
with $ h \equiv \de^{ij} h_{ij}$, the kinetic part $S_K=\int dt d^3
x \sqrt{g}N \frac{2}{\kappa^2}\left(K_{ij}K^{ij}-\f{1}{3}
K^2\right)$ becomes, at the quadratic order,
\begin{\eq}
\label{S_K}
 S_K=\int dt d^3 x \frac{\ep^2}{2
\kappa^2}\left(\dot{h}_{ij} \dot{h}^{ij}-\lambda \dot{h}^2-n_i {\cal
H}^i _{(\ep)} \right),
\end{\eq}
where
\begin{\eq}
\label{mom_const} \f{\ep}{\kappa^2} {\cal H}^i _{(\ep)} \equiv -
\f{2 \ep}{\kappa^2} \left[\pa_t \left( \pa_j h^{ij} -\f{1}{3}
\de^{ij} \pa_j h \right) -\f{1}{3} \pa^i \pa_j n^j -\pa^2 n^i
\right]\approx 0
\end{\eq}
are the momentum constraints at the linear order of $\ep$.

On the other hand, the {\it Diff} (\ref{Diff}) reduces to (see
\ci{Keha,Soti} for comparisons)
\begin{\eq}
\label{Diff:linear}
\de x^i &=&- \ep \xi^i (t, {\bf x}), ~\de t =- \ep g(t), \no \\
\de h_{ij} &=&\pa_i \xi_j +\pa_j \xi_i, \no \\
\de n_i &=&\dot{\xi}_i, ~ \de n =\dot{g}.
\end{\eq}
Here, one can choose, by taking time-independent spatial {\it Diff},
$\xi^i =\xi^i ({\bf x })$,
\begin{\eq}
n_i=0
\end{\eq}
but this does not mean the absence of the momentum constraints $\ep
{\cal H}^i_{\ep}\approx 0$ \ci{Park:0910}. In this case, one can
choose the Ho\v{r}ava's gauge \ci{Hora:08,Hora} for the perturbed
metric $h^{ij}$,
\begin{\eq}
\label{Horava_gauge}
 \pa_j h^{ij} -\f{1}{3} \de^{ij} \pa_j h =0,
\end{\eq}
which is {\it time independent}, according to the momentum
constraints (\ref{mom_const}). Then, the transverse field
\begin{\eq}
\tilde{H}_{ij} \equiv h_{ij} -\f{1}{3} \de_{ij} h,~\pa_i
\tilde{H}_{ij}=0
\end{\eq}
may be introduced. Note that $\tilde{H}_{ij}$ is traceless already,
i.e., ${\tilde{ H}^i}_i=0$.

From these, one obtains
\begin{\eq}
\label{h:decom}
 h_{ij}=\tilde{H}_{ij} + \f{1}{3}
\de_{ij} h
\end{\eq}
with the transverse traceless part $\tilde{H}_{ij}$. Then the
kinetic part (\ref{S_K}), the quadratic order of $\ep$, becomes
\ci{Hora:08,Hora}
\begin{\eq}
\label{S_KH}
 S_K=\frac{\ep^2}{2
\kappa^2} \int dt d^3 x ~
\dot{\tilde{H}}_{ij} \dot{\tilde{H}}^{ij}
.
\end{\eq}
Here, it is important to note that there is no kinetic term for the
scalar mode $h$, as in the $\la=1$ case \ci{Park:0910}. So, there is
no physical excitation for the scalar mode at the conformal point
$\la=1/3$ also.

From the intrinsic curvatures\footnote{I follow the conventions of
Wald \ci{Wald}.} under the perturbations (\ref{pert}),
\begin{\eq}
R^{(3)}_{ij}&=&\f{\ep}{2} \left( \pa^k \pa_i h_{jk} -\pa^2 h_{ij} +
\pa^k \pa_j h_{ik} -\pa_i \pa_j h \right) + R^{(NL)}_{ij} 
, \no \\
R^{(3)}&=&\ep \left( \pa_k \pa_i h^{ik} -\pa^2 h \right)+ {\cal O}
(\ep^2),
\end{\eq}
the potential part which is second order in the (spatial)
derivatives in the flat limit $\La\ra 0$, $S_{V(2)}=\int dt d^3 x
\sqrt{g}N \frac{\kappa^2 \mu^2 \hat{\om}}{8} R^{(3)}$ becomes
\begin{\eq}
\label{S_V2}
 S_{V(2)}=-\frac{\ep^2 \kappa^2 \mu^2
\hat{\om}}{8}\int dt d^3 x \left[ \f{1}{4} h_{ij} \left( -\pa^2
h^{ij} + 2 \pa^k \pa^i {h^j}_k-2 \pa^i \pa^j h + \de^{ij} \pa^2 h
\right)-n {\cal H}^t_{(\ep)} \right],
\end{\eq}
where
\begin{\eq}
\label{ham_const}
 \ep {\cal H}^t _{(\ep)} \equiv - \ep \pa_k (\pa_i
h^{ik}- \pa^k h) \approx 0
\end{\eq}
is the Hamiltonian constraint at the linear order of $\ep$. Here, I
have used \ci{Bogd:0907,Park:0910}
\begin{\eq}
\sqrt{g} R^{(3)}&=&\de^{ij} R_{ij}^{(NL)} +\ep h_{ij} \left(-R^{
ij (L)} +\f{1}{2} \de^{ij} R^{(L)} \right) + {\cal O} (\ep^3) \no \\
&=&\f{\ep}{2} h_{ij} \left(-R^{ij (L) } +\f{1}{2} \de^{ij} R^{(L)}
\right) + {\cal O} (\ep^3),
\end{\eq}
where $R^{(L)}_{ ij}, R^{(NL)}_{ ij}$ denote the linear, non-linear
perturbations of $R^{(3)}_{ij}$, respectively. The action
(\ref{S_V2}), when combined with the Ho\v{r}ava's gauge
(\ref{Horava_gauge}), reduces to
\begin{\eq}
\label{S_V2b}
 S_{V(2)}=\frac{\ep^2 \kappa^2 \mu^2
\hat{\om}}{8}\int dt d^3 x \left[ \f{1}{4} h_{ij} \pa^2 h^{ij}
-\f{5}{36} h \pa^2 h +n {\cal H}^t_{(\ep)} \right].
\end{\eq}

On the other hand, the Hamiltonian constraint (\ref{ham_const}),
when combined with the gauge fixing condition (\ref{Horava_gauge}),
reduces to
\begin{\eq}
\label{ham_const_red}
 {\cal H}^t_{(\ep)}=\f{2}{3} \pa^2 h \approx 0.
\end{\eq}

From the mode decomposition, the second-order spatial derivative
action (\ref{S_V2b}) becomes
\begin{\eq}
 S_{V(2)}=\frac{\ep^2 \kappa^2 \mu^2
\hat{\om}}{8}\int dt d^3 x \left[ \f{1}{4} \tilde{H}_{ij} \pa^2
\tilde{H}^{ij} -\f{2}{36} h \pa^2 h +n {\cal H}^t_{(\ep)} \right].
\end{\eq}
Then the second-order derivative action becomes altogether
\begin{\eq}
 S_{(2)}= \ep^2\int dt d^3 x \left[\frac{1}{2 \kappa^2} \dot{\tilde{H}}_{ij}
\dot{\tilde{H}}^{ij}+ \frac{ \kappa^2 \mu^2 \hat{\om} }{32}
\tilde{H}_{ij} \pa^2 \tilde{H}^{ij}  -\f{2}{9} \f{ \kappa^2 \mu^2
\hat{\om} }{32} h \pa^2 h + \frac{ \kappa^2 \mu^2 \hat{\om}}{16}n
{\cal H}^t_{(\ep)} \right].
\end{\eq}
The first two terms represent the usual transverse traceless
graviton modes $\tilde{H}_{ij}$ with the speed of gravitational
interaction
\begin{\eq}
c_g=\sqrt{ \f{\kappa^4 \mu^2 \hat{\om}}{16}},
\end{\eq}
which agrees with the speed of light $c$ in (\ref{constant}) and
here it is important to note that the propagation can exist due to
the IR modification term with an arbitrary coefficient $\hat{\om}$,
as in the non-conformal cases \ci{Keha,Park:0910}. The next two
terms seem to imply the spatial modulations of the scalar mode $h$
but this mode is completely disappeared in the physical subspace of
the Hamiltonian constraint (\ref{ham_const_red}).

The UV behaviors are governed by the higher derivative terms in
(\ref{SIR}) and the quadratic part of the perturbed action is
\begin{\eq}
\label{S_UV}
 S_{(UV)}= \f{\ep^2}{4} \int dt d^3 x \left[ -\bar{a} \tilde{H}_{ij} \pa^6
\tilde{H}^{ij} + \bar{b} \ep^{ijk} \tilde{H}_{il} \pa^4 \pa_j
{\tilde{H}^l}_{k} +\bar{c} \tilde{H}_{ij} \pa^4 \tilde{H}^{ij} +\f{2
\bar{c}}{27} h \pa^4 h \right],
\end{\eq}
where
\begin{\eq}
\label{detailed}
 \bar{a}=-\f{\kappa^2}{2 \nu^4}, ~ \bar{b}=\f{\kappa^2 \mu}{2
\nu^2},~\bar{c}=-\f{\kappa^2 \mu^2}{8}
\end{\eq}
are the coefficients of $C_{ij} C^{ij},~\epsilon^{ijk}
R^{(3)}_{i\ell} \nabla_{j}R^{(3)\ell}{}_k$, and $(R^{(3)}_{ij}
R^{(3)ij}-\f{1}{3} R^{(3)} R^{(3)})$, respectively. The first three
terms provide the modified dispersion relation $\om^2 \sim k^6+
\cdots$ for the transverse traceless modes.\footnote{From the parity
violating term, $\bar{b} \ep^{ijk} \tilde{H}_{il} \pa^4 \pa_j
{\tilde{H}^l}_{k}$, as in the non-conformal cases $\la \neq 1/3$,
the asymmetry in the right and left-handed circular polarization
modes $\tilde{H}_{R/L}=\f{1}{\sqrt{2}}(\tilde{H}_{11} \pm i
\tilde{H}_{12})$ propagating along $x^3$ direction
$(\tilde{H}_{3i}=0)$, with the (full) dispersion relations
$\om^2_{R/L}=\f{\ka^4 \mu^2 \hat{\om}}{16} k^2_{R/L}-\f{\bar{c}
\ka^2}{2} k^4_{R/L} \pm \f{\bar{b} \ka^2}{2} k^5_{R/L}-\f{\bar{a}
\ka^2}{2} k_{R/L}^6$, is also generated such as the gravitational
waves are {\it chiral} \ci{Taka:0904,Bogd:0907}. This is in contrast
to the naive expectation from the non-conformal case, which gives
the suppression of the chiral modes in the $\la \ra 1/3$ limit.}
Here, the (UV) detailed balance with the particular values of the
coefficients (\ref{detailed}) do not have any role. The last term
contains higher spatial derivatives of the scalar mode $h$ but this
does not appear in the physical subspace again. Here, the
non-existence of sixth derivative terms for the scalar mode is the
result of the detailed balance in sixth order,
\begin{\eq}
C_{ij} C^{ij} = \al \nabla_{i}R^{(3)}_{jk} \nabla^{i}{R^{(3)}}^{jk}
+\be \nabla_{i}R^{(3)}_{jk}\nabla^{j} {R^{(3)}}^{ik}
+\ga\nabla_{i}R^{(3)}\nabla^{i}R^{(3)}
\end{\eq}
with $\al=1,\be=-1,\ga=-1/8$. On the other hand, for arbitrary
values of $\al,\be,\ga$ one obtains
\begin{\eq}
C_{ij} C^{ij} = -\f{\al \ep^2 }{4} \tilde{H}_{ij} \pa^6
\tilde{H}^{ij} +\f{\ep^2 }{4} \left(\f{2}{3}\right)^2 \left( \f{3
\al}{2} +\be +4 \ga \right) h \pa^6 h
\end{\eq}
and there are sixth derivative terms for the scalar mode $h$. But,
even in this case, these terms do not produce the propagation in the
physical subspace.

\section{Concluding Remarks}

In conclusion, I have constructed the Ho\v{r}ava model for the
$\la=1/3$ case, where the local anisotropic Weyl symmetry exists in
the UV limit, in addition to the foliation-preserving {\it Diff},
such as the lapse function $N$ is a non-projectable function on the
foliation generally. By considering the linear perturbations around
the Minkowski vacuum background, I showed that the scalar graviton
mode is completely disappeared in the physical spectrum, as
expected, and only the usual tensor graviton modes remain. This
situation is analogous to $\la=1$, which is Lorentz invariant in the
IR limit.

The existence of the UV conformal symmetry is unique to the theory
with the detailed balance. This may explain the importance of the
detailed balance in quantum gravity. Generally, the parameter
$\lambda$ would run in RG flows due to quantum corrections except
for the fixed point where high symmetries occur. In this context, it
would be quite probable that $\la=1/3$ be the UV fixed point since
there is no UV symmetry enhancement outside of $\la=1/3$. However,
it is not known yet whether this symmetry is high enough for the
fixed point or not. It would be a challenging problem to check this
explicitly in RG flows as in $\la=1$, which seems to be the IR fixed
point, similarly.

{\it Note added}: After the appearance of this paper, a related
papers which considered the black hole solutions appeared
\ci{Kiri:0910,Capa:0911}. Later at the fully non-linear orders for
the IR limit of Ho\v{r}ava gravity (\ref{lEH}), it has been also
found that the number of physical degrees of freedom is the same as
GR, which implies that there is no non-perturbative generation of
scalar graviton as well, in the IR limit. \ci{Bell}

\section*{Acknowledgments}

This work was supported by the Korea Research Foundation Grant
funded by Korea Government(MOEHRD) (KRF-2010-359-C00009).

\newcommand{\J}[4]{#1 {\bf #2} #3 (#4)}
\newcommand{\andJ}[3]{{\bf #1} (#2) #3}
\newcommand{\AP}{Ann. Phys. (N.Y.)}
\newcommand{\MPL}{Mod. Phys. Lett.}
\newcommand{\NP}{Nucl. Phys.}
\newcommand{\PL}{Phys. Lett.}
\newcommand{\PR}{Phys. Rev. D}
\newcommand{\PRL}{Phys. Rev. Lett.}
\newcommand{\PTP}{Prog. Theor. Phys.}
\newcommand{\hep}[1]{ hep-th/{#1}}
\newcommand{\hepp}[1]{ hep-ph/{#1}}
\newcommand{\hepg}[1]{ gr-qc/{#1}}
\newcommand{\bi}{ \bibitem}


\begin{thebibliography}{999}


\bibitem{Hora:08}
  P.~Horava,
  ``Membranes at Quantum Criticality,''
  JHEP {\bf 0903}, 020 (2009)
  [arXiv:0812.4287 [hep-th]].

\bibitem{Hora} P.~Horava,
  ``Quantum Gravity at a Lifshitz Point,''
  Phys.\ Rev.\  D {\bf 79}, 084008 (2009)
  [arXiv:0901.3775 [hep-th]].

\bi{Cai:0905} R.~G.~Cai, B.~Hu and H.~B.~Zhang,
  ``Dynamical Scalar Degree of Freedom in Horava-Lifshitz Gravity,''
  Phys.\ Rev.\  D {\bf 80}, 041501 (2009)
  [arXiv:0905.0255 [hep-th]].

\bibitem{Keha}
  A.~Kehagias and K.~Sfetsos,
  ``The black hole and FRW geometries of non-relativistic
  gravity,'' Phys. Lett. B {\bf 678}, 123 (2009) [arXiv:0905.0477 [hep-th]].

\bi{Chen:0905_2} B.~Chen, S.~Pi and J.~Z.~Tang,
  ``Scale Invariant Power Spectrum in Ho\v{r}ava-Lifshitz Cosmology without
  Matter,'' JCAP {\bf 0908}, 007 (2009)
  [arXiv:0905.2300 [hep-th]].

\bi{Char} C.~Charmousis, G.~Niz, A.~Padilla and P.~M.~Saffin,
  ``Strong coupling in Horava gravity,'' JHEP {\bf 0908}, 070 (2009)
  [arXiv:0905.2579 [hep-th]].

\bi{Li} M.~Li and Y.~Pang,
  ``A Trouble with Ho\v{r}ava-Lifshitz Gravity,'' JHEP {\bf 0908},
  015 (2009)
  [arXiv:0905.2751 [hep-th]].

\bibitem{Soti}
  T.~Sotiriou, M.~Visser and S.~Weinfurtner,
  ``Phenomenologically viable Lorentz-violating quantum gravity,''
  Phys. Rev. Lett. {\bf 102}, 251601 (2009)
  [arXiv:0904.4464 [hep-th]];
  ``Quantum gravity without Lorentz invariance,'' JHEP {\bf 0910}, 033 (2009)
  [arXiv:0905.2798 [hep-th]].

\bibitem{Kim}
  Y.~W.~Kim, H.~W.~Lee and Y.~S.~Myung,
  ``Nonpropagation of scalar in the deformed Ho\v{r}ava-Lifshitz
  gravity,'' Phys.\ Lett.\  B {\bf 682}, 246 (2009)
 [arXiv:0905.3423 [hep-th]].

\bibitem{Muko:0905}
  S.~Mukohyama,
  Phys.\ Rev.\  D {\bf 80}, 064005 (2009)
  [arXiv:0905.3563 [hep-th]].

\bibitem{Gao:0905}
  X.~Gao, Y.~Wang, R.~Brandenberger and A.~Riotto,
  ``Cosmological Perturbations in Ho\v{r}ava-Lifshitz Gravity,''
  [Phys.\ Rev.\  D {\bf 81}, 083508 (2010)]
[arXiv:0905.3821 [hep-th]].

\bibitem{Blas:0906}
  D.~Blas, O.~Pujolas and S.~Sibiryakov,
  ``On the Extra Mode and Inconsistency of Horava Gravity,'' JHEP {\bf 0910}, 029 (2009)
[arXiv:0906.3046 [hep-th]].

\bibitem{Koba:0906}
  A.~Kobakhidze,
  ``On the infrared limit of Horava's gravity with the global Hamiltonian
  constraint,''
  Phys.\ Rev.\  D {\bf 82}, 064011 (2010)
[arXiv:0906.5401 [hep-th]].

\bibitem{Bogd:0907}
  C.~Bogdanos and E.~N.~Saridakis,
  ``Perturbative instabilities in Horava gravity,'' Class.\ Quant.\ Grav.\  {\bf 27}, 075005 (2010)
[arXiv:0907.1636 [hep-th]].

\bibitem{Wang:0907}
  A.~Wang and R.~Maartens,
  ``Linear perturbations of cosmological models in the Horava-Lifshitz theory
  of gravity without detailed balance,'' Phys.\ Rev.\  D {\bf 81}, 024009 (2010)
[arXiv:0907.1748 [hep-th]].

\bibitem{Park:0910}
  M.~I.~Park,
  ``Remarks on the Scalar Graviton Decoupling and Consistency of Ho\v{r}ava
  Gravity,'' Class.\ Quant.\ Grav.\  {\bf 28}, 015004 (2011)
 [arXiv:0910.1917 [hep-th]].

\bibitem{Park:0905}
  M.~I.~Park,
  ``The Black Hole and Cosmological Solutions in IR modified Horava
  Gravity,'' JHEP {\bf 0909}, 123 (2009)
  [arXiv:0905.4480 [hep-th]].

\bibitem{Park:0906}
  M.~I.~Park,
  ``A Test of Horava Gravity: The Dark Energy,'' JCAP {\bf 1001}, 001 (2010)
[arXiv:0906.4275 [hep-th]].

\bibitem{Gong:1002}
  J.~O.~Gong, S.~Koh and M.~Sasaki,
  ``A complete analysis of linear cosmological perturbations in
  Ho\v{r}ava-Lifshitz gravity,''
  Phys.\ Rev.\  D {\bf 81}, 084053 (2010)
  [arXiv:1002.1429 [hep-th]].

\bibitem{Park:0705}
  M.~I.~Park,
  ``Holography in Three-dimensional Kerr-de Sitter Space with a   Gravitational
  Chern-Simons Term,''
  Class.\ Quant.\ Grav.\  {\bf 25}, 135003 (2008)
  [arXiv:0705.4381 [hep-th]].

\bibitem{Witt}
  E.~Witten,
  ``(2+1)-Dimensional Gravity as an Exactly Soluble System,''
  Nucl.\ Phys.\  B {\bf 311}, 46 (1988).

\bibitem{Hora:0909}
  P.~Horava and C.~M.~Melby-Thompson,
  ``Anisotropic Conformal Infinity,''
  arXiv:0909.3841 [hep-th].

\bi{Wald} R. M. Wald, ``General Relativity" (University of Chicago
Press., Chicago, 1984).

\bibitem{Taka:0904}
  T.~Takahashi and J.~Soda,
  ``Chiral Primordial Gravitational Waves from a Lifshitz Point,''
  Phys.\ Rev.\ Lett.\  {\bf 102}, 231301 (2009)
  [arXiv:0904.0554 [hep-th]].

\bibitem{Kiri:0910}
  E.~Kiritsis and G.~Kofinas,
  ``On Horava-Lifshitz 'Black Holes','' JHEP {\bf 1001}, 122 (2010)
  [arXiv:0910.5487 [hep-th]].

\bibitem{Capa:0911}
  D.~Capasso and A.~P.~Polychronakos,
  ``General static spherically symmetric solutions in Horava
  gravity,'' Phys.\ Rev.\  D {\bf 81}, 084009 (2010)
[arXiv:0911.1535 [hep-th]].

\bibitem{Bell}
  J.~Bellorin and A.~Restuccia,
  ``On the consistency of the Horava Theory,''
  arXiv:1004.0055 [hep-th].

\end{thebibliography}
\end{document}